\documentclass[12pt]{article}
\usepackage[margin=2cm]{geometry}
\usepackage{comment}
\usepackage{amsmath,amssymb,extarrows,mathtools,graphicx,subfigure,setspace}
\usepackage{cite}
\usepackage{slashed}
\usepackage{color}
\makeatother

\newcommand{\be}{\begin{equation}}
\newcommand{\bea}{\begin{eqnarray}}
\newcommand{\eea}{\end{eqnarray}}
\newcommand{\ba}{\begin{array}}
\newcommand{\ea}{\end{array}}
\newcommand{\ee}{\end{equation}}
\newcommand{\bes}{\begin{equation*}}
\newcommand{\beas}{\begin{eqnarray*}}
\newcommand{\eeas}{\end{eqnarray*}}
\newcommand{\bas}{\begin{array*}}
\newcommand{\eas}{\end{array*}}
\newcommand{\ees}{\end{equation*}}

\numberwithin{equation}{section}
\begin{document}
\onehalfspacing
\vfill
\begin{titlepage}
\vspace{10mm}
\begin{flushright}
 IPM/P-2012/050 \\
FPAUO-12/12\\
\end{flushright}
  %
\vspace*{20mm}
\begin{center}
{\Large {\bf  Charged Black Branes  with Hyperscaling Violating Factor     }\\
}

\vspace*{15mm}
\vspace*{1mm}
{Mohsen Alishahiha$^{a}$, Eoin \'O Colg\'ain$^{b}$ and  Hossein Yavartanoo$^{c}$ }

 \vspace*{1cm}

{\it ${}^a$ School of Physics, Institute for Research in Fundamental Sciences (IPM)\\
P.O. Box 19395-5531, Tehran, Iran \\ }
 \vspace*{0.5cm}
{\it ${}^b$ Departamento de F\'isica, Universidad de Oviedo, 33007 Oviedo, Spain \\ }
 \vspace*{0.5cm}
{\it ${}^c$ Department of Physics, Kyung-Hee University, Seoul 130-701, Korea }

\vspace*{2cm}
\end{center}

\begin{abstract}
We present an  analytic solution of a charged black hole with hyperscaling violating factor in 
an Einstein-Maxwell-Dilaton model where the scalar potential is key to the existence of a solution. This solution provides a candidate gravitational description of theories with hyperscaling violation at both finite temperature and finite charge density.
Using this background we explore certain features of  these theories via AdS/CFT correspondence. Finally, we discuss embeddings based on the well-known sphere reductions of ten and eleven-dimensional supergravity.

\end{abstract}

\end{titlepage}

\section{Introduction}
Gauge/gravity duality\cite{Maldacena:1997}  may be thought of as a practical tool to study strongly-coupled systems near critical points where the system exhibits a scaling symmetry. Generally at
a critical point the system may be described by a conformal field theory (CFT). From the gauge/gravity 
point of view this means that the gravitational theory is defined on a metric which is asymptotically
locally Anti-de Sitter (AdS). 

On the other hand, in many physical systems  critical points are governed
by dynamical scalings in which, even though the system  exhibits a scaling symmetry, space and time 
scale differently under this symmetry. A prototype example of such critical points is a Lifshitz fixed point 
where the system is spatially isotropic and scale invariant, though there is an anisotropic scaling in the time direction characterized by a dynamical exponent, $z$. More precisely the system is invariant 
under the following scale symmetry
\be
r\rightarrow \lambda^z r,\;\;\;\;\;\;\;\;\;\;x_i\rightarrow \lambda x_i,
\ee
where $t$ is time and $x_i$ are spatial coordinates of a $d$ dimensional space (in our case the
$d$ dimensional space is $R^d$).

In the framework of gauge/gravity duality it is then important to find  gravitational theories that provide
a gravity description of  Lifshitz fixed points.  Indeed, such a gravity dual for the Lifshitz fixed point was introduced in \cite{KLM} where the authors considered a gravitational theory which admits a 
solution exhibiting the above scaling symmetry (see also \cite{Koroteev:2007yp}
for earlier work on a geometry with Lifshitz scaling.). The corresponding metric may be written
as
\be
ds^2=-r^{2z}dt^2+r^2 \sum_{i=1}^d dx_i^2+\frac{dr^2}{r^2}.
\ee

In terms of \textit{bottom-up} model building, a natural set-up to consider is Einstein-Maxwell-Dilaton (EMD) theory, a setting in which numerous studies of 
Lifshitz-like black brane geometries have already appeared\cite{ Goldstein:2009cv, Cadoni:2009xm,Kiritsis, Eric,  GKPTIW, Peet, BBPZ, Cadoni:2011kv, Sandip, Berglund,Tarrio:2011de, Myung:2012cb,Pal:2012zn}. 
As is well appreciated at this stage, with the inclusion of  dilaton and Abelian gauge fields, it is possible to find even more sophisticated metrics generalizing Lifshitz  which, on top of an anisotropic scaling, have also an overall hyperscaling
factor. More precisely, one may have a geometry in the form of\cite{Kiritsis}
\be
\label{metric1}
ds^2=r^{\frac{-2\theta}{d}} \left(-r^{2z}dt^2+r^2\sum_{i=1}^ddx_i^2
+\frac{dr^2}{r^2} \right),
\ee
where the constants $z$ and $\theta$ are  dynamical and  hyperscaling violation exponents, respectively. This geometry has been the subject of considerable recent attention and hyperscaling violating Lifshitz solutions have been found in a host of different settings \cite{Singh:2010zs, Narayan:2012hk, Singh:2012un, Dey:2012tg, Dey:2012rs, Dey:2012fi}. This is the most general geometry which is spatially homogeneous and covariant under the following scale transformations
\be
t\rightarrow \lambda^z t, \quad r\rightarrow \lambda^{-1} r, \quad x_i\rightarrow \lambda x_i, \quad  
ds_{d+2} \rightarrow \lambda^{\frac{\theta}{d}} ds_{d+2}.
\ee
Note that with a non-zero $\theta$, the distance is not
invariant under the scaling which in the context of AdS/CFT indicates  violations of 
hyperscaling in the dual field theory. More precisely, while in $(d+1)$-dimensional theories 
without  hyperscaling,  the entropy  
scales as $T^{d/z}$ with temperature in the present case where the metric exhibits hyperscaling, it scales
as $T^{(d-\theta)/z}$ \cite{{Gouteraux:2011ce},{Huijse:2011ef}}. 

An interesting feature of the above metric is that for the special value of the hyperscaling violation
exponent $\theta=d-1$, the holographic entanglement entropy\cite{RT:2006PRL,RT:2006} exhibits
a logarithmic violation of area law\cite{{Ogawa:2011bz},{Huijse:2011ef}, Dey:2012hf}, indicating that the background \eqref{metric1} could provide
a gravitational dual for a theory with  an ${\cal O}(N^2)$ Fermi surface, where $N$ is the number of 
degrees of freedom, or alternatively the number of colors for an $SU(N)$ gauge theory. 

These observations indicate that backgrounds whose asymptotic behavior coincides with the above metric may be of interest to condensed 
matter physics, thus making it natural to further explore gauge/gravity duality for these backgrounds, some of the aspects of which have already been touched upon in \cite{{Dong:2012se},{Alishahiha:2012cm}}.  As the eventual goal of this program is to make contact with condensed matter, a natural generalisation involves backgrounds with finite charge density. Recently, we have witnessed the first construction of finite temperature and finite charge density solutions in a top-down construction with probe D-branes \cite{Ammon:2012je}. Here, as a further step in this direction, we identify what we believe to be the first back-reacted hyperscaling violating backgrounds with non-zero temperature and charge. This extends hyperscaling geometries in EMD theories at finite temperature \cite{{Huijse:2011ef},{Dong:2012se}, {Cadoni:2012uf}} to finite charge density. 

This article is organized as follows. In the next section we will find hyperscaling 
violating charged black brane solutions in an EMD theory with a
non-trivial potential for the dilaton. We observe that the presence and the form of the 
potential is crucial to have  the solution. We shall also show that at zero temperature where the solution 
is extremal, the near-horizon geometry develops an $AdS_2$ factor indicating that 
theories with hyperscaling violation  at finite charge density  and zero temperature may be described by an 
IR fixed point. In section three we study holographic entanglement entropy for a hyperscaling violating 
black brane solution, where we explore how the non-zero charge affects the entanglement 
entropy.  In section four we probe the background by a charged  fermion where we investigate
the possibility of having an ${\cal O}(N^0)$ Fermi surface in the model.
In section five we study optical conductivity for theories with hyperscaling violation.  The last section is 
devoted to discussions, in particular, those related to candidate embeddings. \\

\textbf{Note added:} While we were writing up we became aware of further Lifshitz solutions with hyperscaling violation \cite{pet}.


\section{Gravity solution}

In this section we introduce our  Einstein-Maxwell-Dilaton
model and illustrate hyperscaling violating charged black branes for this theory.  For the asymptotically Lifshitz geometry an instructive analytic solution for charged black branes 
has been found \cite{Tarrio:2011de}. To find such a solution one needs at least two $U(1)$ gauge
fields coupled to a scalar field. The first $U(1)$ together with the scalar field is necessary to 
generate an anisotropic scaling, while the second one is required to have a charged solution.
It is also possible to find asymptotically charged black holes as well. For this 
purpose one should add another $U(1)$ gauge field to be responsible for the 
compacted horizon\cite{Tarrio:2011de}. 

To find  hyperscaling violating charged black branes we will closely  follow the approach of
\cite{Tarrio:2011de}. We note, however, that in order to get a non-trivial hyperscaling factor, 
besides the action considered in \cite{Tarrio:2011de}, the model considered here has to have a non-trivial potential 
for the  scalar field as well. In what follows we only consider  the model  with two $U(1)$ gauge fields and therefore we will find  black branes charged under one $U(1)$ 
gauge  field. We expect that it is straightforward to generalize for black holes with more gauge fields.

To proceed we consider a minimal model as follows
\be
\label{action}
S=-\frac{1}{16\pi G}\int d^{d+2}x\sqrt{-g}\left[R-\frac{1}{2}(\partial\phi)^2+V(\phi)-\frac{1}{4}
\sum_{1=1}^2 e^{\lambda_i\phi}F_i^2\right], 
\ee
where we have not confined ourselves to any particular dimensionality. 
For the potential of  the scalar field, motivated by the typical exponential potentials of string theory, we will consider the 
following potential\footnote{Note that we do not expect to find this potential in string theory truncations of AdS vacua, but those of domain walls, i.e. the near-horizon of various branes. } 
\be
V=V_0e^{\gamma\phi}.
\ee
Here $\lambda_1, \lambda_2, \gamma$ and $V_0$ are free parameters of the model. 
We will see  that with this simple potential, the model does, indeed, admit solutions with 
hyperscaling violation.

The equations of motion of the above  action read
\bea
&&R_{\mu\nu}+\frac{V(\phi)}{d}g_{\mu\nu}=\frac{1}{2}\partial_\mu\phi\partial_\nu\phi
+\frac{1}{2}\sum_{i=1}^2e^{\lambda_i\phi}\left(F_{i\;\mu}^\rho F_{i\;\rho\nu}-\frac{g_{\mu\nu}}{2d}F_i^2\right),\cr &&\cr
&& \nabla^2\phi=-\frac{dV(\phi)}{d\phi}+\frac{1}{4}\sum_{i=1}^2\lambda_i e^{\lambda_i\phi} 
F_i^2,\;\;\;\;\;\;
\nabla_\mu\left(\sqrt{-g}e^{\lambda_i\phi}F_i^{\mu\nu}\right)=0.
\eea
Let us consider the following ansatz for the metric, scalar and gauge field
\be\label{ansatz}
ds^2=r^{2\alpha}\left(-r^{2z}f(r)dt^2+\frac{dr^2}{r^2f(r)}+r^2d\vec{x}^2\right),\;\;\;\;\phi=\phi(r),
\;\;\;\;\; F_{i\;rt}\neq 0,
\ee
and assume that the other components of gauge fields are set to zero.

From the Maxwell equations of motion, using the above ansatz, one finds
\be
F_{i\; rt}=e^{-\lambda_i\phi}r^{\alpha(2-d)+z-d-1}\rho_i.
\ee 
Then by combining the  $tt$ and $rr$ components of the Einstein equation one has
\be\label{tr}
R^t_t-R^r_r=-\frac{1}{2}g^{rr}(\partial_r\phi)^2.
\ee
On the other hand, by calculating the Ricci tensor for our ansatz we get
\be
R^t_t-R^r_r=-d(\alpha+1)(\alpha+z-1)r^{-2\alpha} f(r).
\ee
Plugging this expression back into  the equation \eqref{tr}  one can determine scalar field as follows
\be\label{sol-phi}
e^\phi=e^{\phi_0} r^{\sqrt{2d(\alpha+1)(\alpha+z-1)}}=e^{\phi_0} r^\beta.
\ee
Note that in order to have a well defined solution one has to assume $(\alpha+1)(\alpha+z-1)\geq 0$.
Indeed it can be seen that this assumption is a consequence of the null energy condition. More
precisely consider a null vector as\cite{Tarrio:2011de} $\xi^\mu=(\sqrt{g^{rr}},\sqrt{g^{tt}},0)$, then 
\be
T_{\mu\nu}\xi^\mu\xi^\nu\sim R^r_r-R^t_t=d(\alpha+1)(\alpha+z-1)r^{-2\alpha} f(r) \geq 0,
\ee 
which in turn translates into $(\alpha+1)(\alpha+z-1)\geq 0$.

To find the metric one utilizes the other components of the Einstein equations. Indeed for our ansatz one has
\be
R^x_x=-(\alpha+1) r^{-\alpha(d+2)-z-d+1}\left(r^{d(\alpha+1)+z}f(r)\right)',
\ee
where, as in standard practice, a prime represents derivative with respect to $r$. Therefore the $xx$ component of the 
Einstein equations of motion reads
\be
\left(r^{d(\alpha+1)+z}f(r)\right)'=\frac{r^{\alpha(d+2)+z+d-1}}{(\alpha+1)}\left(\frac{V_0e^{\gamma\phi}}{d}-\frac{1}{2d}\sum_{i=1}^2 e^{-\lambda_i\phi}\rho_i^2
r^{-2d(\alpha+1)}\right). 
\ee
Using the solution of $\phi$ \eqref{sol-phi}  one arrives at
\be
\left(r^{d(\alpha+1)+z}f(r)\right)'=\frac{r^{\alpha(d+2)+z+d-1}}{(\alpha+1)}\left(\frac{V_0e^{\gamma\phi_0} r^{\gamma\beta}}{d}
-\frac{1}{2d}\sum_{i=1}^2 e^{-\lambda_i\phi_0}\rho_i^2
r^{-2d(\alpha+1)-\lambda_i\beta}\right), 
\ee
\newpage
which can be integrated to find the function $f$ of the metric as follows
\bea
f(r)&=&-mr^{-d\alpha-z-d}+\frac{V_0e^{\gamma\phi_0} r^{\gamma\beta+2\alpha}}{d(\alpha+1)
(\gamma\beta+\alpha(d+2)+z+d)}\cr &&\cr
&&-\sum_{i=1}^2\rho_i^2e^{-\lambda_i\phi_0}\frac{r^{-2\alpha(d-1)-\beta\lambda_i-2d)}}
{2d(\alpha+1)(\alpha(2-d)+z-d-\beta\lambda_i)},
\eea
where $m$ is a constant which can be related to the mass of the black brane.

 With this expression, we have
found all functions involved in the ansatz \eqref{ansatz}. The resulting solution has seven
free parameters: $\alpha, z, m, \phi_0, \rho_1, \rho_2$ and the radius of the metric which  has 
already been set to unity. On the other hand the action has four free parameters
$\lambda_1,\lambda_2,\gamma$ and $V_0$.  Therefore four parameters of the solution can be fixed 
by the four parameters of the model yielding a solution with three parameters which
correspond to the mass, charge and the value of the scalar field, let's say, at horizon. 

Note that so far we have not used all equations of motion. Indeed, in order, to find 
the parameters of the solution in terms of those in the action,  we  now utilize  the remaining  equations of motion. 
To proceed, we note that in order to maintain the asymptotic behavior of the hyperscaling 
violating metric, as it is evident from the expression of $f$,  one must fix $\gamma$ as follows
\be
\gamma=-\frac{2\alpha}{\beta}.
\ee
From this expression it is clear that to find a charged black brane (hole) with hyperscaling
violating factor, it is crucial to have a non-trivial potential. 

From  the equation of motion of the scalar field,  we get
\be
\left(\frac{4\beta }{d(\alpha+1)}-\frac{8\alpha}{\beta}\right)V_0e^{-2\alpha\phi/\beta}=
\sum_{i=1}^2 e^{\lambda_i\phi} F_i^2\left(\lambda_i-\frac{\beta}{d(\alpha+1)}\right)
\ee
This equation can  be solved for parameters $\lambda_1,\lambda_2$ and $\rho_1$. Actually this 
equation may be solved in different, but rather equivalent,  ways. In particular one may find\footnote{
It is worth mentioning that instead of finding the parameters of the solution in terms of those
in action, we have fixed the action in terms of  parameters of the solution.}
\be
\lambda_1= -\frac{2\alpha(d-1)+2d}{\sqrt{2d(\alpha+1)(\alpha+z-1)}},\;\;\;\lambda_2=
\sqrt{\frac{2(\alpha+z-1)}{d(\alpha+1)}},\;\;\;
\rho_1^2=\frac{2V_0(z-1)e^{-\sqrt{\frac{2d(\alpha+1)}{\alpha+z-1}}\phi_0}}{d\alpha+d+z-1}.
\ee 
With these expressions, $\rho_2$ remains as an undetermined free parameter which, indeed, can be
identified with the charge of the solution. The last parameter, $V_0$, may also be fixed by setting 
the constant term in the expression of $f$ to one. Doing so, one arrives at
\be
{V_0}=e^{\frac{2 \alpha  {\phi_0}}{\sqrt{2d (1+\alpha ) (-1+z+\alpha )}}} 
(d \alpha+z+d-1 ) (d\alpha+z+d  ).
\ee
With this equation we have fixed all four free parameters, though there are still unused equations
of motion. It is, then, important to check whether the other equations hold without imposing any further 
constraints on the parameters of the solution. In particular one of the non-trivial equations need to be checked is
the $tt$ component of the Einstein equations of motion. Indeed it is easy to see that this 
equation is also satisfied without imposing any further constraints. 

Therefore, to summarize,  we note that the action \eqref{action} admits hyperscaling violating charged black 
brane solutions which can be recast to the  following form\footnote
{Note that  in order to follow the standard notation in the literature we set $\alpha=-\theta/d$,
 where $\theta\geq 0$ is
the hyperscaling violation exponent. Observe also that this solution is not valid for 
$\theta=d$ where $\alpha=-1$.}
\bea\label{solution}
ds^2&=&r^{-2\frac{\theta}{d}}\left(-r^{2z}f(r)dt^2+\frac{dr^2}{r^2f(r)}+r^2d\vec{x}^2\right),\cr &&\cr
 F_{1\;rt}&=&\sqrt{2(z-1)(z+d-\theta)}e^{\frac{\theta(1-d)/d+d}{\sqrt{2(d-\theta)(z-1-\theta/d)}}\phi_0}
\;r^{d+z-\theta-1},\cr &&\cr
F_{2\;rt}&=&Q\sqrt{2(d-\theta)(z-\theta+d-2)} e^{-\sqrt{\frac{z-1-\theta/d}{2(d-\theta)}}\;\phi_0}\;r^{-(z+d-\theta-1)},\cr &&\cr
e^{\phi}&=&e^{\phi_0}r^{\sqrt{2(d-\theta)(z-1-\theta/d)}},
\eea
with\footnote{As a consistency check , we note that for $d+z-\theta=2$, this solution reduces to an uncharged 
solution obtained in\cite{Dong:2012se}.}
\be
f(r)=1-\frac{m}{r^{z+d-\theta}}+\frac{Q^2}{r^{2(z+d-\theta-1)}}.
\ee
This is indeed a charged black brane solution whose radius of horizion, $r_H$,  is obtained by setting 
$f=0$ which leads to the following algebraic equation for $r_H$:
\be
r_H^{2(d+z-\theta-1)}-m r_{H}^{d+z-\theta-2}+Q^2=0.
\ee
The corresponding Hawking temperature is also found
\be
T=\frac{(d+z-\theta)r_H^z}{4\pi}\left(1-\frac{(d+z-\theta-2)Q^2}{d+z-\theta}r_H^{2(\theta-d-z+1)}\right).
\ee
Therefore the extremal limit is given by 
 \be
r_H^{2(d+z-\theta-1)}=\frac{(d+z-\theta-2)}{d+z-\theta}Q^2,
 \ee
in which case the function $f$ reads
\be
f=1-\frac{2(d+z-\theta-1)}{d+z-\theta-2}\left(\frac{r_H}{r}\right)^{d+z-\theta}+
\frac{d+z-\theta}{d+z-\theta-2}\left(\frac{r_H}{r}\right)^{2(d+z-\theta-1)}.
\ee
In this limit it is easy to see that the solution near the horizon develops an $AdS_2\times R^{d-1}$
geometry. More precisely, in the extremal case one can consider the following change of coordinates
\be
r-r_H=\frac{\epsilon r_H^2}{(d+z-\theta)(d+z-\theta-1)\zeta},\;\;\;\;\;\;\;\;\;\;\;t=\frac{\tau}{\epsilon
r_H^z}.
\ee
Then the near-horizon limit is defined by the limit  $\epsilon\rightarrow 0$ for which the solution
\eqref{solution} reads
\bea\label{exsolution}
ds^2&=&r_H^{2-2\frac{\theta}{d}}\left(\frac{-d\tau^2+d\zeta^2}{(d+z-\theta)(d+z-\theta-1)\zeta^2}+d\vec{x}^2\right),\cr &&\cr
 A_{1\tau}&=&\frac{\sqrt{2(z-1)(z+d-\theta)}}{(d+z-\theta)(d+z-\theta-1)}\;\frac{e^{\frac{\theta(1-d)/d+d}{\sqrt{2(d-\theta)(z-1-\theta/d)}}\phi_0}}{r_H^{\theta-d-1}}\;\frac{1}{\zeta},\cr &&\cr
A_{2\tau}&=&\frac{Q\sqrt{2(d-\theta)(z-\theta+d-2)}}{(d+z-\theta)(d+z-\theta-1)}\; \frac{e^{-\sqrt{\frac{z-1-\theta/d}{2(d-\theta)}}\;\phi_0}}{r_H^{2z+d-\theta-3}}\;\frac{1}{\zeta},\cr &&\cr
e^{\phi}&=&e^{\phi_0}r_H^{\sqrt{2(d-\theta)(z-1-\theta/d)}},
\eea
which is $ AdS_2\times R^d$ with two charges.

Following the general idea of gauge/gravity duality we would like to consider the  solution 
\eqref{solution} as a gravitational background dual to a theory with 
hyperscaling violation at finite temperature and finite charge density.  We note that 
the gravitational description of theories with hyperscaling violation at finite temperature 
has already been studied in \cite{Dong:2012se}. Actually the aim of  the following sections is to 
further explore the holographic description of the theories with hyperscaling violation when the effect
of charge has also been taken into account.  

As an observation, we note that  due to non-zero charge, the black brane could be extremal with 
near horizon of $AdS_2$ as we just demonstrated. It means that at low temperature the theory
flows to an IR fixed point, where the theory may be governed by a two-dimensional CFT.


\section{Entanglement entropy}
The aim of this section is to study the holographic entanglement entropy of a hyperscaling 
violating charged black brane by making use of the AdS/CFT correspondence. To compute the entanglement entropy via AdS/CFT correspondence
one needs to minimize a surface in the bulk gravity.  More precisely, given a gravitational  theory with
 the  bulk Newton's constant $G_N$, the holographic entanglement entropy 
is given by \cite{RT:2006PRL,RT:2006}
\be\label{EE}
S_A=\frac{\mathrm{Area}(\gamma_A)}{4G_N},
\ee
where $\gamma_A$ is the  minimal surface in the bulk whose boundary coincides with the boundary of the entangling region. 

In this section we will only consider the extremal case where the temperature is zero
and therefore we will be able to study the effects of non-zero  charge on the entanglement entropy.
In the non-extremal case where we have non-zero temperature, the results essentially reduce to 
those in\cite{Dong:2012se}. This is due to the fact that, in this case, the charge effects are subleading 
to the thermal behavior of the entanglement entropy.

To proceed, let is consider a long strip in the dual theory given by
\be
-\frac{\ell}{2}\leq x_1\leq \frac{\ell}{2},\;\;\;\;\;\;\;0\leq x_i\leq L\;\;\;\;\;\;{\rm for}\;i=2,\cdots,d.
\ee
The surface $\gamma_A$ is defined by a hypersurface whose boundary coincides with the above strip and
has a profile in the bulk of the metric \eqref{exsolution} given by $x_1=x(r)$. The induced metric 
on this hypersurface is 
\be
ds_{\rm in}^2=\rho^{2\frac{\theta}{d}-2}\left[\left(\frac{1}{f(\rho)}+x'(\rho)^2\right)d\rho^2
+dx_i^2\right],
\ee
where we have made a change of coordinate, $r=\frac{1}{\rho}$,  and prime  now represents  the derivative
with respect to $\rho$. Therefore the area of the 
hypersurface reads
\be
A=L^{d-1}\int d\rho \;\rho^{\theta-d}\sqrt{f^{-1}+x'^2}.
\ee
As is well known, the above expression for the area may be treated as an action for a
one dimensional mechanical system where the momentum conjugate of the field $x$ is 
conserved. So that
\be
\frac{x'}{\sqrt{f^{-1}+x'^2}}=\left(\frac{\rho_0}{\rho}\right)^{\theta-d},
\ee
with $\rho_0$ representing the turning point where $x'(\rho_0)\rightarrow \infty$. It is then 
easy to find the width of the strip as follows
\be
\ell=2\rho_0\int_0^1\frac{\xi^{d-\theta}f^{-1/2}(\xi,\rho_0/\rho_H)}
{\sqrt{1-\xi^{2(d-\theta)}}},
\ee
where 
\be
f(\xi,\frac{\rho_0}{\rho_H})
=1-\frac{2(d+z-\theta-1)}{d+z-\theta-2}\left(\frac{\rho_0}{\rho_H}\right)^{d+z-\theta}\xi^{d+z-\theta}+
\frac{d+z-\theta}{d+z-\theta-2}\left(\frac{\rho_0}{\rho_H}\right)^{2(d+z-\theta-1)}
\xi^{2(d+z-\theta-1)}.
\ee
The area is then given by
\be
A=L^{d-1}\rho_0^{\theta-d+1}\int_{\frac{\epsilon}{\rho_0}}^1\frac{\xi^{\theta-d}f^{-1/2}(\xi,\rho_0/\rho_H)}
{\sqrt{1-\xi^{2(d-\theta)}}}. 
\ee
Now the aim is  to perform the above integrals and  eliminate $\rho_0$ to find the entanglement 
entropy as a function of the width of the strip $\ell$.  Of course in general it cannot be done analytically,
though in special limits it it possible to do so.

Let's assume that the width of strip is very small, which in turn results in the turning point being much smaller than the radius of the horizon, {\it i.e.} $\rho_0\ll \rho_H$.  For this case when $\theta\neq d-1$, at leading order, the finite part of the
entanglement entropy is
\be
S_{\rm finite}\approx \frac{ L^{d-1}\ell^{\theta-d+1}}{4G_N}\left(1+a_0(z,d,\theta) Q^{\frac{d+z-\theta}
{d+z-\theta-1}}\ell^{d+z-\theta}+a_1(z,d,\theta) Q^2\ell^{2(d+z-\theta-1)}\right),
\ee
while for $\theta= d-1$ the entanglement entropy exhibits a logarithmic violation 
of area law
\be
S\approx \frac{L^{d-1}}{4G_N}\ln\frac{\ell}{\epsilon}+c_0(z,d)Q^{\frac{z+1}{z}}\frac{L^{d-1}\ell^{z+1}}
{4 G_N\epsilon^{z+1}}+c_1(d,z) Q^2\frac{L^{d-1}\ell^{2z}}{4G_N\epsilon^{2z}}.
\ee
 This is in fact a signature that the dual theory has an ${\cal O}(N^2)$ Fermi surface which can be
probed by the enanglement entropy\cite{{Ogawa:2011bz},{Huijse:2011ef}}.  Here $a_i(z,d,\theta)$ and
$c_i(z,d)$ are numerical constants.

On the other hand if we want to consider a strip with large width, one should let the hypersurface 
extend all the way to the deep IR region, {\it i.e.} $\rho_0\sim\rho_H$. In this case, at leading 
order, one arrives at
\be
S_{\rm finite}\approx\frac{L^{d-1}\ell}{4G_N}\;Q^{\frac{d-\theta}{d+z-\theta-1}}.
\ee
It is worth mentioning that if we compute the Hawking entropy of the extremal charged black brane
 the resultant  entropy is
\be
S_{\rm H}\sim\frac{{\rm Vol}_d}{4G_n}\;Q^{\frac{d-\theta}{d+z-\theta-1}},
\ee
which has the same behavior as the above entanglement entropy. Note that here ${\rm Vol}_d$ is the
volume of $d$ dimensional space generated by $\vec{x}$.


\section{Charged Fermion probe}

In the previous section we have studied holographic  entanglement entropy for theories
with hyperscaling violation. If the theory  has an ${\cal O}(N^2)$ Fermi surface, the
holographic entanglement entropy is able to probe it. Indeed, in this case the entanglement
entropy exhibits a logarithmic violation of the area law. Actually for the model we are studying
when $\theta=d-1$ with or without charge, we have a logarithmic violation of area law, showing 
that it has an ${\cal O}(N^2)$ Fermi surface (see also \cite{{Ogawa:2011bz},{Huijse:2011ef}}). 

On the other hand when $\theta\neq d-1$, the entanglement entropy has a power law behavior
which indicates that the dual theory does not have an ${\cal O}(N^2)$ Fermi surface. 
Nevertheless one may wonder whether the model still has an ${\cal O}(N^0)$ Fermi surface.
Of course in this case the entanglement entropy cannot probe the Fermi surface. In fact, 
to see whether the system exhibits an ${\cal O}(N^0)$ Fermi surface one could 
probe the system by a charged fermion. 
Then to see a hint of a Fermi surface, one should look for a sharp behavior in the fermionic retarded Green's function at finite momentum and small frequencies\cite{{Lee:2008xf},{Liu:2009dm},{Cubrovic:2009ye},{Faulkner:2009wj},{Faulkner:2010zz},{Hartnoll:2011dm},{Cubrovic:2011xm}} (for a review
see\cite{Faulkner:2011tm}).

In this section we will probe  the extremal  hyperscaling violating 
black brane geometry  by a charged fermion\footnote{Fermions in asymptotically Lifshitz
geometries have been sudied in 
\cite{{Korovin:2011kw},{Gursoy:2011gz},{Alishahiha:2012nm},{Wu:2012fk},{Gursoy:2012ie}}.}.
This can be used to read off the retarded Green's function of a fermionic operator in the dual theory,  
which is a theory with hyperscaling violation at finite density. Using the pole structure of the
corresponding retarded Green's function, we gain an insight into whether, or not, a Fermi surface exists. 
Since we will have to deal with fermions, for simplicity, we will only consider the four
dimensional ($d=2$) case. It is, of course, straightforward to generalize our analysis here to arbitrary dimensions.

To proceed, let us consider  a four dimensional charged Dirac fermion on the background 
\eqref{solution} in the extremal limit
\be S_{\rm bulk} =\int d^4x \sqrt{-g}\,i\bar{\Psi} \,\left[\frac{1}{2}\left(\Gamma^a\!\stackrel{\rightarrow}D_a-
\stackrel{\leftarrow}D_a\!\Gamma^a\right)-m\right]\Psi\label{Action}.
\ee
Here
$\Gamma^a D_a=(e_\mu)^a\Gamma^\mu[\partial_a+\frac{1}{4}(\omega_{\rho\sigma})_a\Gamma^{\rho\sigma}-iq A_{2\;a}]
$, with
$\Gamma^{\mu\nu}=\frac{1}{2}[\Gamma^\mu,\Gamma^\nu]$.

It is important to note that in order to find  the equations of motion for this action, one needs 
to use the variational principle, which for a space with a boundary, always comes with a proper boundary condition. Of course, the  boundary term is not necessarily  unique and, indeed,  for the above
fermion action it was shown in\cite{Laia:2011zn}  that there are numerous ways to make use of different boundary terms to make the variational principle well-defined. Of course, a given boundary term
may break the symmetries of the model. In what follows we will consider the standard
boundary condition (in the notation of \cite{Laia:2011zn} ). The equation of motion\footnote{As it is evident 
fron our solution, the background is charged under the second gauge field and the first gauge is 
present to  provide an anisotropic scaling along the time direction. Therefore in what follows we consider a 
 fermion which is charged under the second gauge field.}
\be
\left((e_\mu)^a\Gamma^\mu\left[\partial_a+\frac{1}{4}(\omega_{\rho\sigma})_a\Gamma^{\rho\sigma}-iqA_{2\;a}\right]-m\right)\Psi=0.
\ee
Here the non-zero components of the vielbeins and spin connections for the  metric in equation
\eqref{solution} are 
\be
(e_t)^a=\frac{r^{\theta/2-z}}{\sqrt{f}}\delta^a_t,\;\;\;\;\;\;(e_r)^a=r^{\theta/2+1}\sqrt{f}\delta^a_t,
\;\;\;\;\;\; (e_i)^a=r^{\theta/2-1}\delta^a_i,
\ee
and 
\be
(\omega_{tr})_a=-\frac{(r^{2z-\theta})'}{2r^(z-1-\theta)}\delta_{at},\;\;\;\;\;\;\;\;
(\omega_{ir})_a=\frac{2-\theta}{2}r\sqrt{f}\delta_{ai}.
\ee
Then through the choice of $\Psi=(-g g^{rr})^{-1/4}e^{-i\omega t+ik.x}\psi(r)$ with $g$ being the determinant of the metric, the above equation of motion reduces to\footnote{Here we have set the radius of the horizon to one.}
\be\label{EOM}
\bigg[rf^{1/2}\Gamma^r\partial_r-\frac{i}{r^zf^{1/2}}\left(\omega+q \mu_2(1-\frac{1}{r^{z-\theta}})\right)
\Gamma^t+\frac{i }{r} \Gamma\cdot k-\frac{m}{r^{\theta/2}}\bigg]\psi(r)=0,
\ee
where $\mu_2=Q\sqrt{\frac{2(2-\theta)}{z-\theta}}\;e^{-\sqrt{\frac{2z-2-\theta}{4(z-\theta)}}\phi_0}$.
Note that we have normalized the gauge field in  such a way that it vanishes at the horizon.  

To find the retarded Green's function, we will utilize a numerical method to solve the equation of
motion. To proceed, it is useful to consider  the following representation for the four dimensional gamma matrices
\begin{eqnarray}\label{basis2}
\Gamma^r=
\begin{pmatrix}
-\sigma^3&0\\0&-\sigma^3
\end{pmatrix},
\Gamma^t=
\begin{pmatrix}
i\sigma^1&0\\0&i\sigma^1
\end{pmatrix},
\Gamma^1=
\begin{pmatrix}
-\sigma^2&0\\0&\sigma^2
\end{pmatrix},
\Gamma^2=
\begin{pmatrix}
0&-i\sigma^2\\i\sigma^2&0
\end{pmatrix}.
\end{eqnarray}
Due to rotational symmetry in the spatial directions we may set $k_2=0$. Then using
the notation
\begin{eqnarray}
\psi=
\begin{pmatrix}
\Phi_1\\ \Phi_2
\end{pmatrix},
\end{eqnarray}
the equation of motion \eqref{EOM} reduces to the following decoupled equations
\be\label{Eq}
\bigg[rf^{1/2}\partial_r-\frac{1}{r^{z}f^{1/2}}\left(\omega+q \mu_{2}(1-\frac{1}{r^{z-\theta}})\right)
i\sigma^2+\frac{m}{r^{\theta/2}}\sigma^3-(-1)^\alpha\frac{k_1}{r} \sigma^1\bigg]\Phi_\alpha=0,
\ee
for $\alpha=1,2$.  It is easy to see that
\be
\Phi_\alpha\sim a_\alpha \begin{pmatrix} 0\\ 1\end{pmatrix}+
 b_\alpha \begin{pmatrix} 1\\ 0\end{pmatrix},\;\;\;\;\;\;\;\;\;{\rm for}\;r\rightarrow
\infty.
\ee
Note  that the asymptotic behavior of different components of  the fermions are independent
of $m$. This is in contrast to the Lifshitz case where we have\cite{Alishahiha:2012nm}
\be
\Phi_\alpha\sim a_\alpha r^m \begin{pmatrix} 0\\ 1\end{pmatrix}+
 b_\alpha  r^{-m}\begin{pmatrix} 1\\ 0\end{pmatrix},\;\;\;\;\;\;\;\;\;{\rm for}\;r\rightarrow
\infty.
\ee
Actually, this means that in our case we can use both standard and alternative quantization to read the 
retarded Green's function. In other words, one may consider either of $a_\alpha$ or $b_\alpha$ 
as the source and the other as the response. 

To find the retarded Green's function, following \cite{Faulkner:2009wj}, it is useful
to set $\chi_1=\psi_1/\psi_2$ and $\chi_2=\psi_3/\psi_4$ where
$\psi_i$'s are defined via $\Phi_1=(\psi_1, \psi_2), \Phi_2=(\psi_3,\psi_4)$.
These parameters satisfy the
following equations
\bea
&&{rf^{1/2}}\partial_r\chi_1+\frac{2m}{r^{\theta/2}}\chi_1-\left(\frac{\Omega}{r^zf^{1/2}}+
\frac{k_1}{r} \right) \chi_1^2=\frac{\Omega}{r^zf^{1/2}}-\frac{k_1}{r} ,\cr &&\cr
&&{rf^{1/2}}\partial_r\chi_2+\frac{2m}{r^{\theta/2}}\chi_2-\left(\frac{\Omega}{r^zf^{1/2}}-
\frac{k_1}{r} \right) \chi_2^2=\frac{\Omega}{r^zf^{1/2}}+\frac{k_1}{r} ,
\eea
where
\be
\Omega=\omega+q \mu_2 (1-\frac{1}{r^{z-\theta}}).
\ee
Using these equations the  retarded Green's function is essentially given in terms
of functions $G_1(k,\omega)$ and $G_2(k,\omega)$ where\footnote{Here we set $k_1=k$.}
\be
G_\alpha(k,\omega)=\lim_{r\rightarrow \infty} \chi_\alpha,\;\;\;\;\;\;\;\;\;\;
{\rm for}\;\alpha=1,2.
\ee
with the 
ingoing boundary condition imposed at the horizon,  which in our notation is  \cite{Faulkner:2009wj}
\be
\chi_\alpha|_{\rm horizon}=i.
\ee
More concretely,  the corresponding retarded Green's function may be given by
\be
G(k,\omega)= -
\begin{pmatrix}
G_1(k,\omega)&0\\0&G_2(k,\omega)
\end{pmatrix}.
\ee
Therefore the spectral function reads
\be
{\cal A}(k,\omega)=\frac{1}{\pi}{\rm Im} \bigg(G_1(k,\omega)+G_2(k,\omega)\bigg).
\ee
Having found expressions for the retarded Green's function, it is an easy task to further determine its behavior as a function of $k$ and $\omega$. 
We have illustrated the behavior in figure 1 for  $m=0, q\mu_2=\sqrt{3}$. 

We have plotted the spectral function for $\theta=1, z=2$ and $\theta=0.4, z=1.4$.
To compare the results with those in the RN AdS case studied in \cite{Faulkner:2009wj},  we have also plotted 
this case as well. Note that in all cases we have $z-\theta=1$. Observe that since the retarded Green's function is an even function of $k$  we only considered $k > 0$.
\begin{figure}
\begin{center}
\includegraphics[height=4.5cm, width=6cm]{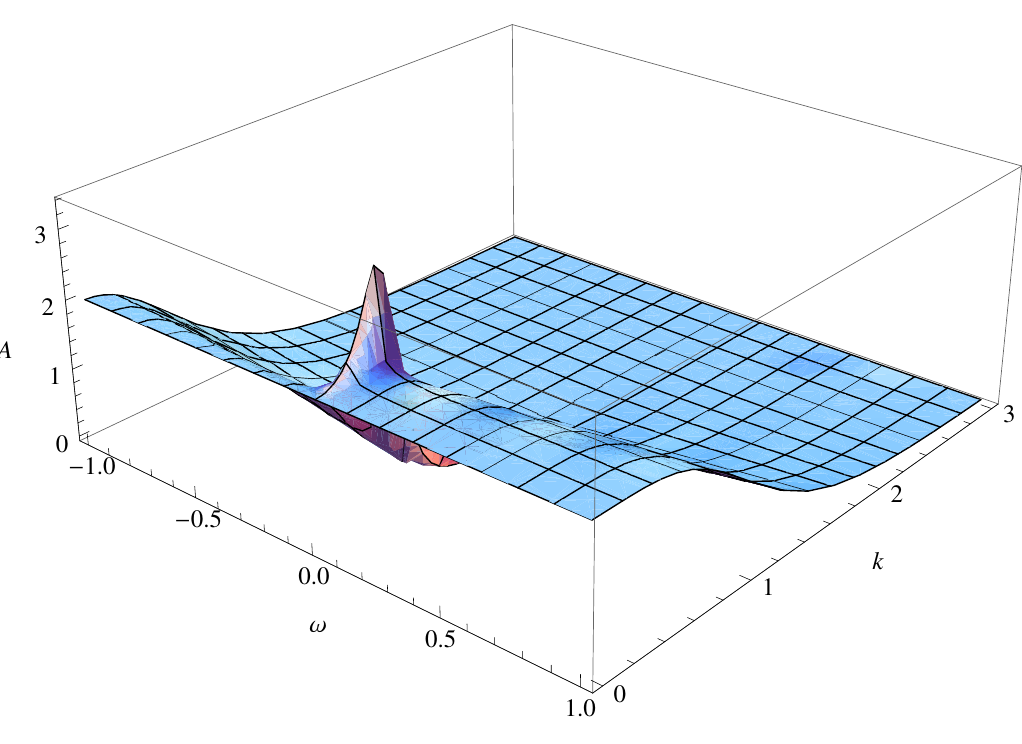}
\hspace{1cm}
\includegraphics[height=4.5cm, width=6cm]{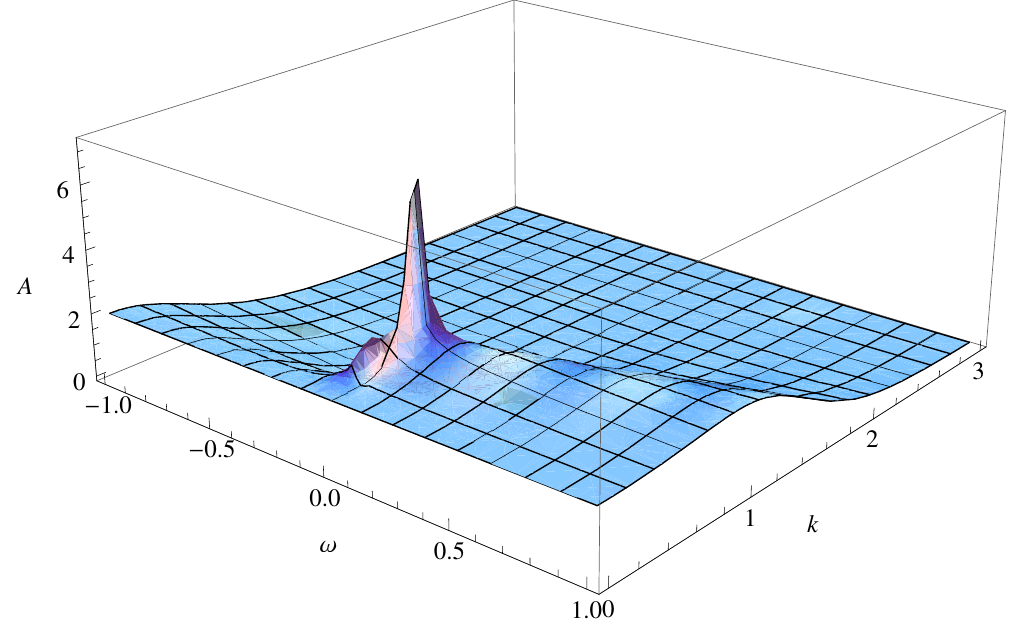}
\hspace{1cm}
\includegraphics[height=4.5cm, width=6cm]{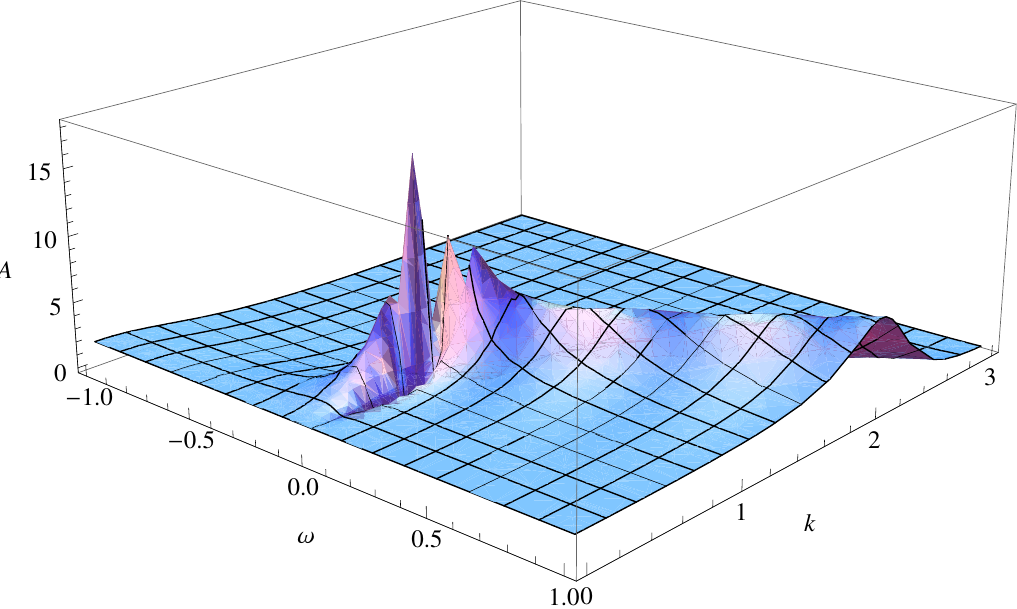}
\caption{The behavior of spectral functions the for hyperscaling violating geometry with $
\theta=1,z=2$ (left), and $\theta=0.4, z=1.4$ (right). To compare the results with that of RN AdS black 
brane we have also plotted this case which in our notation corresponds to $\theta=0,z=1$ (see
\cite{Faulkner:2009wj}). }\label{fig5}
\end{center}
\end{figure}

One observes that the effect of  increasing $\theta$ or $z$, is to broaden the sharp peak representing the 
${\cal O}(N^0)$ Fermi surface. Moreover, as we see, even though for  $\theta=0.4, ~z=1.4$ the
entanglement entropy does not exhibit a violation of area law, at the probe limit there is almost
a sharp peak in the fermionic retarded Green's function which might indicate that the system
has a Fermi surface. 

For $ z=2$  and $\theta=1$, where the entanglement entropy exhibits a violation of 
area law indicating a Fermi surface at ${\cal O}(N^2)$, we still have a peak in the spectral function
around $\omega\approx 0$,  though it is not as sharp as that in the RN AdS black brane.

\section{Optical conductivity}

In this section we study the optical conductivity of a finite density system whose gravity dual is given by \eqref{solution}. The local $U(1)$ gauge field in the bulk is dual to a conserved current of a global $U(1)$ symmetry in the dual theory. In terms of the conserved current, the optical conductivity can be obtained from the Kubo formula as follows
\bea\label{Kubo}
\sigma(\omega)=\frac{1}{i\omega}\langle J(\omega)J(-\omega)\rangle_{\rm retarded},
\eea
where $J(\omega)$ is the conserved current of the $U(1)$ global symmetry evaluated at zero spatial momentum. The right hand side can be calculated using the  AdS/CFT correspondence. Since we are working in the classical gravity regime, the leading contribution of the classical fluctuations of the gauge field is of ${\cal O}(N^2)$ which is due to the black brane background. In particular, it is then interesting to determine 
the optical conductivity for the case of $\theta=d-1$ where we have an ${\cal O}(N^2)$ Fermi
surface.


To proceed, we consider small fluctuations of the gauge field in the $x_1$-direction, $\delta A_{x_1}\equiv a_1$. In general, turning on small fluctuations of the gauge field causes a back-reaction on the other fields of the model,  so we have to solve the equations of motion for all the fluctuations.  More precisely one may have
\bea
A_{2\;\mu}\rightarrow A_{2\;\mu} \delta_{\mu 0}+a_\mu, \hspace{1cm}g_{\mu\nu}\rightarrow g_{\mu\nu}+h_{\mu\nu},\hspace{1cm}\phi\rightarrow\phi+\delta\phi.
\eea
Since we are interested in calculating the two point function, it is enough to expand the action up to quadratic level in the fluctuations. To do so, it is convenient to first use the gauge freedom and set $h_{r\mu}=a_r=0$. Moreover at the level we are interested in we may set  $\delta\phi=0$.
With these assumptions it is straightforward to  compute the  quadratic terms of the action from which
we can read off the equations of motion for the fluctuations.
For our purpose, we will focus on the zero momentum case which means we may set $a_1=a(r)e^{-i\omega t}$. In this case the gauge field fluctuations mix only with the $h_{tx}$ component of the metric fluctuations. Therefore we find two coupled differential equations. Nevertheless one may eliminate $h_{tx}$ from the equations leading to the following differential equation for the gauge fluctuation:\footnote{For more details see for example \cite{Alishahiha:2012ad} and
also  Appendix 6.B of \cite{Iqbal}.}
\bea\label{Ax}
\partial_r\left[\sqrt{-g}e^{\lambda_2\phi}g^{rr}g^{xx}a'(r)\right]+
\left(\frac{{\cal Q}^2 g_{rr}g_{tt}}{\sqrt{-g}g_{xx}}-\omega^2\sqrt{-g}g^{tt}g^{xx} e^{\lambda_2\phi}\right)a(r)=0,
\eea
where ${\cal Q}=\sqrt{-g}e^{\lambda_2\phi} g^{tt} g^{rr}F_{2\;rt}$. For the solution
\eqref{solution} the above equation reads\footnote{Note that in what follows, for simplicity, we have set $\phi_0=0$.}
\be
\partial_r[r^{3(z-1)+d-\theta}f a'(r)]-\left[2Q^2(d-\theta)(z-\theta+d-2)r^{z-3+\theta-d}-
\frac{\omega^2 r^{z-\theta+d-5}}{f}\right]a(r)=0.
\ee
Then for the special case of $\theta=d-1$ the equation reduces to
\be\label{gfeq}
\partial_r[r^{3z-2}f a'(r)]-\left[2Q^2(z-1)r^{z-4}-
\frac{\omega^2 r^{z-4}}{f}\right]a(r)=0,\;\;{\rm with}\;f=1-(1+\frac{Q^2}{r_H^{2z}})\left(\frac{r_H}{r}
\right)^{z+1}+\frac{Q^2}{r^{2z}}.
\ee
It is interesting to note that the dependence on $\theta$ and $d$ drop out from the equation and it simply depends on a  given $z$.  

Now the aim is to solve the above equation to find a solution with an ingoing boundary condition at the horizon. Then the corresponding retarded Green's function can
 be read from the asymptotic behavior of the solution near the boundary. Indeed one finds
\be 
a\sim A+\frac{B}{r^{3(z-1)}},
\ee
for $r\rightarrow \infty$ and at the horizon we get
\be 
a\sim (r-r_h)^{\pm i\frac{\omega^2 r_H^{z}}{(z+1)r_H^{2z}-Q^2(z-1)}}.
\ee
Therefore with a proper choice of the solution near the horizon (ingoing solution) the 
retarded Green's function is given by
\be
\langle J_1(\omega)J_1(-\omega)\rangle_{\rm retarded}\sim \frac{B}{A}.
\ee

In order to find the retarded Green's function we utilize numerical methods. To proceed, we note that the equation \eqref{gfeq} has four free parameters, $Q$, $\omega, z$ and $r_H$. In fact setting $z=2$, and
using ``NDSolve" in Mathematica the real and imaginary parts of the conductivity as functions of 
frequency at a fixed temperature
may be found as depicted in figure 2.
\begin{figure}
\begin{center}
\includegraphics[scale=.8]{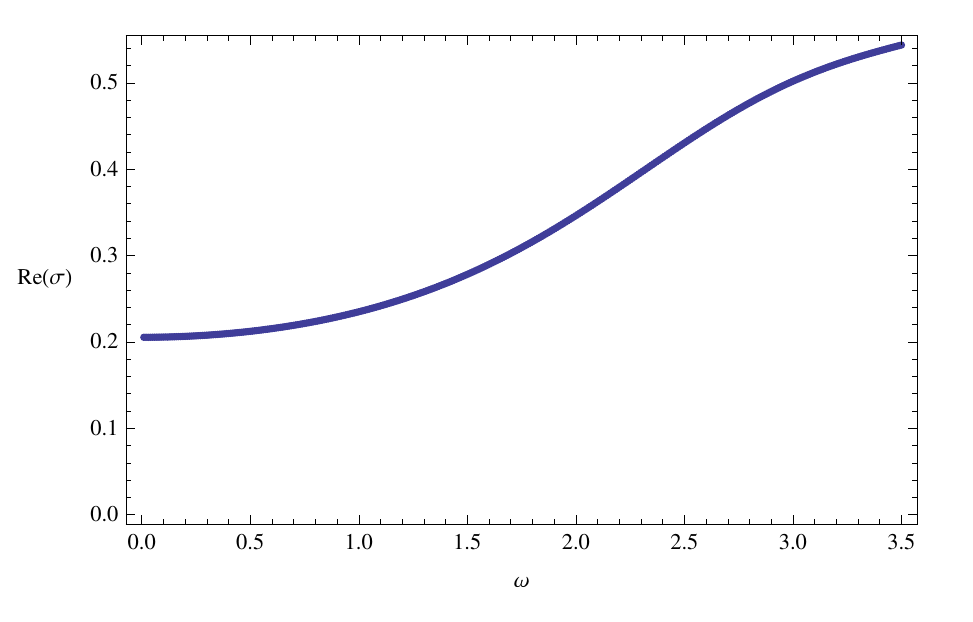}\label{sigma}
\hspace{1cm}
\includegraphics[scale=.8]{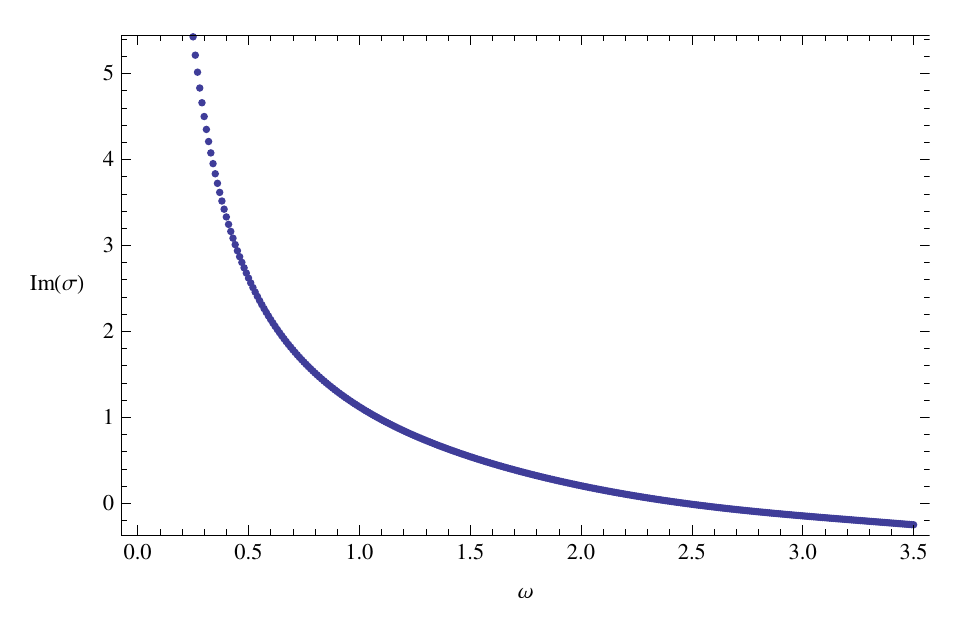}
\caption{The real and imaginary parts of the conductivity versus $\omega$ for a fixed temperature.
To find this plot we have set $Q=\sqrt{3}$, and  $r_H=1.25$.}
\end{center}
\end{figure}

The real part of the conductivity has the typical shape and from the imaginary part of the
conductivity we observe that there is a delta function behavior in the real part.
 It is worth noting that the delta function behavior is an artifact of simplifications in the gravity calculation which should be compared with a sample without impurities. Adding impurities would broaden the delta function into a Drude peak. From the gravity point of view, this can be done by imposing a non-trivial boundary condition on the source of an operator at the boundary so that the translational invariance is broken. In this case, the Drude peak will appear from gravity calculations as well \cite{Tong:2012}.

On the other hand, for a fixed $\omega$ one can find the behavior of the conductivity as a function of 
temperature. In particular, setting $\omega=0.001$,  the real part of the 
conductivity can be found numerically as shown in figure 3.
\begin{figure}
\begin{center}
\includegraphics[scale=.8]{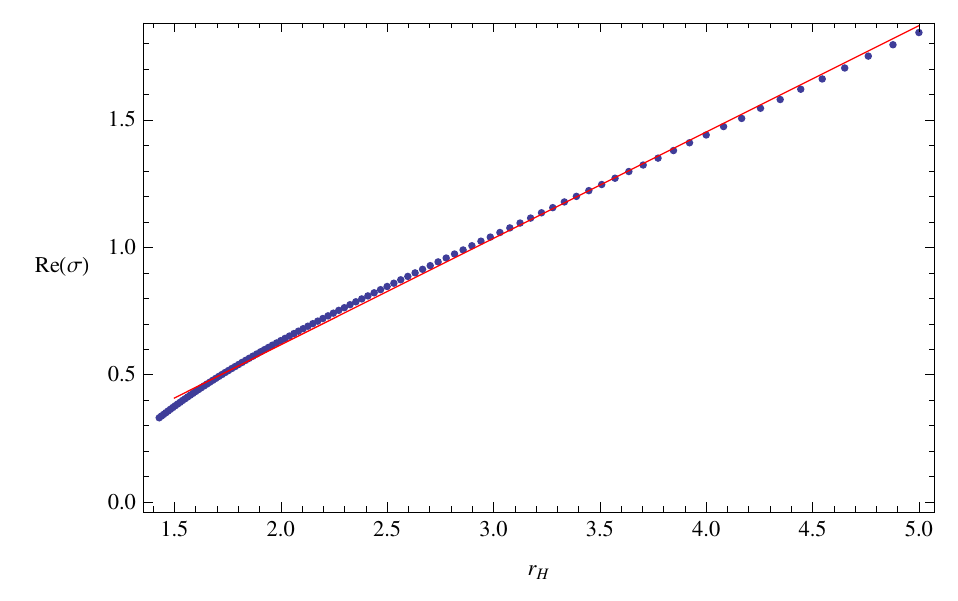}\label{sigma}
\caption{The real part of the conductivity versus $r_H$ for a fixed frequency.
To find this plot we have set $Q=\sqrt{3}$, and  $\omega=0.001$. The numerical solution is shown by dots,
while the best fit is drawn by a red line.}
\end{center}
\end{figure}

Again using numerical methods the best fit one finds for the real part of the conductivity is
\be
{\rm Re}(\sigma)\approx - 0.22 + 0.42 r_H.
\ee
Of course one may use the expression of the Hawking tempereture to write
the result in terms of $T$. Indeed, doing so, one finds $ {\rm Re}(\sigma)\approx T^{1/2}$.

\section{Discussions} 
In this paper we have considered an EMD theory with a non-trivial potential 
for the dilaton. We have found an analytic solution representing a hyperscaling
violating charged black brane solution. To do so, we have observed that the minimal model
should contain  at least two gauge fields, one to produce an anisotropic scaling and the other 
to correspond to a charge. 
It is known that to generate the required Lifshitz anisotropy, the first gauge field always diverges at the boundary,
while the second falls off as a normal gauge field. Although in this paper we have only considered black branes with only one charge,  we stress that generalizations to black holes with more charges are expected to be straightforward\cite{Tarrio:2011de}.

Following the general thrust of the gauge/gravity philosophy, the resulting background \eqref{solution} may provide gravity duals for strongly coupled theories with 
hyperscaling violation at finite charge density  and temperature. It is then natural to explore 
different aspects of the theory by making use of gauge/gravity duality. Note that in order 
to study the effect of the charge, we have only considered the extremal case.
Of course, one could have repeated the analysis for non-zero temperature, though the results  at leading order
would be the same as those obtained in \cite{Dong:2012se}. This is because the effects associated to the charge are 
subleading. This can be seen, for example, in the expression of the temperature when
expressed in terms of radius of the horizon. Moreover, the entropy of the black branes \eqref{solution}
at leading order in charge is given by
\be
S\sim T^{\frac{d-\theta}{z}}\left[1+\frac{(d-\theta)(d+z-\theta-2)}{z(d+z-\theta)}\;Q^2\;
T^{\frac{2(\theta+1-d-z)}{z}}\right].
\ee
An immediate observation is that in order to have positive specific heat, at leading order we find the constraint $d\geq \theta$\cite{Dong:2012se}, which was our assumption throughout the
present paper.

We have used the extremal solution \eqref{exsolution} to compute the entanglement entropy
for a strip and we have been able to obtain the leading order correction to the entanglement entropy due to 
non-zero charge. We have observed that, even with non-zero charge, when $\theta=d-1$,
the entanglement entropy, at leading order, exhibits a logarithmic violation of area law 
indicating that the dual theory may have an ${\cal O}(N^2)$ Fermi surface\footnote{Another interesting 
observation is the charge dependence of the corrections is non-analytic.}. This, of course, would not
exclude the possibility of having a Fermi surface of the order of one for arbitrary values of $\theta$. 

Indeed, in order to study an order one Fermi surface we have also to probe the extremal background 
with a charged fermion.  In the process, we have shown the retarded Green's function of the dual fermionic
operator may have a pole in terms of a momentum for $\omega\approx 0$ indicating that 
there is the Fermi surface. We note, however, that as one increases $\theta$ or $z$, the sharp peak
becomes smooth and eventually, for large enough $\theta$ and/or $z$, the peak disappears, suggesting 
that the system does not have a Fermi surface.

We have also studied optical conductivity at ${\cal O}(N^2)$ order numerically for the case of 
$\theta=d-1$ and $z=2$. We have found that in this case the real part of the conductivity, at leading
order in charge, goes as $T^{1/2}$. Indeed, we could have done it for arbitrary $\theta, d$ and $z$.
In this case one finds
\be
{\rm Re}(\sigma)\sim T^{\frac{d-\theta}{z}}.
\ee

As we have already mentioned, at zero temperature the geometry will develop an $AdS_2$ geometry
with a non-zero gauge field at the near-horizon limit. This indicates that at low energy the theory
is described by a two-dimensional CFT, which is dual to the $AdS_2$ geometry. However, 
due to the non-zero gauge field, the theory becomes unstable for sufficiently large charges. This is due 
to the fact that the scaling dimensions of operators in the dual theory become imaginary.  If the 
operator is bosonic we would have bose condensation, while for a fermionic operator it may back-react to modify the geometry.  Basically,  
the situation is very similar to that in a RN black hole \cite{Faulkner:2011tm}.  

It is also important to note that due to the hyperscaling violation and the  behavior of the dilaton field 
at large distances, the background \eqref{metric1} cannot provide a dual description
of a theory in all range of energies from UV to IR \cite{{Bhattacharya:2012zu},{Kundu:2012jn}}
(see also\cite{Dong:2012se}) . In fact, gravity on the background \eqref{solution} may be considered 
as an effective theory which is valid over an intermediate energy scale. 

What about string theory embeddings? We respond by echoing the sentiment of others by saying that it would be interesting to see if one can embed our solution to string theory. Here, to highlight that we do not expect this to be easy, we add some comments about embeddings based on the well-known sphere-reductions of ten and eleven-dimensional supergravity. In particular, we will focus on the truncations to $U(1)^2$, $U(1)^3$ and $U(1)^4$ gauged supergravities without axions which exist in in seven, five and four-dimensions respectively \cite{Cvetic:1999xp}. Starting with $d=5$ in our notation, neglecting the obvious drawback of it having a relatively more complicated potential, the $U(1)^2$ seven-dimensional gauged supergravity cannot be truncated further to a single scalar while retaining two independent gauge fields, so we can quickly eliminate this case.  

The potentials for $d=2, 3$ are both of the form 
\be
V(\phi) = V_1 e^{\gamma_1 \phi} + V_2 e^{\gamma_2 \phi},  
\ee
and further truncations to a single scalar and two gauge fields can be found. So now we can return to section 2 and generalize the potential by adding the required extra exponential term. However, for these embeddings, if we require that one of the $\rho_i$ of our ansatz is a free parameter, such that it corresponds to the charge of a black hole, we find that the only solution is  $\alpha=-1$, which we have already excluded from our solutions. 

So, the simplest truncations based on sphere-reductions do not allow for embeddings. Given that we initially only considered an exponential potential, it seems promising that one could embed this using a domain wall type set-up, since such simple potentials typically arise there. However, we still require the exact field content of a single scalar and two gauge fields. We leave further exploration to future work.

\vspace*{1cm}

\section*{Acknowledgments}
We are grateful to Patrick Meessen and Nilanjan Sircar for discussion on related topics. M. A would like to thank TH-divsion of CERN where this work has done for very warm hospitality.
 The research of M.A. is supported by Iran National Science Foundation (INSF). 
 E. \'O C is partially supported by the research grants MICINN-09-FPA2009-07122 and MEC-DGI-CSD2007-00042.
The work of H.Y  is supported by the National Research Foundation of Korea Grant funded by the Korean Government (NRF-2011- 0023230).

\end{document}